\documentclass[a4paper,twocolumn,american,pra,aps,showpacs]{revtex4}
\usepackage{graphicx}
\usepackage{amssymb}

\makeatletter

\providecommand{\tabularnewline}{\\}

\usepackage{graphicx}

\usepackage{bm}

\usepackage{babel}
\makeatother
\begin{document}

\title{Perfect quantum state transfer with randomly coupled quantum chains}

\author{Daniel Burgarth and Sougato Bose}
\affiliation{Department of Physics \& Astronomy, University College London, Gower
St., London WC1E 6BT, UK}

\begin{abstract}
We suggest a scheme that allows arbitrarily perfect state transfer
even in the presence of random fluctuations in the couplings of a
quantum chain. The scheme performs well for both spatially correlated
and uncorrelated fluctuations if they are relatively weak (say 5\%).
Furthermore, we show that given a quite arbitrary pair of quantum
chains, one can check whether it is capable of perfect transfer by
only local operations at the ends of the chains, and the system in
the middle being a {}``black box''. We argue that unless some specific
symmetries are present in the system, it \emph{will} be capable of
perfect transfer when used with dual-rail encoding. Therefore our
scheme puts minimal demand not only on the control of the chains when
using them, but also on the design when building them.
\end{abstract}


\maketitle

\section{Introduction}

Recently, much interest has been devoted to quantum communication
with quantum chains \cite{SB03,JOSEPH,key-7,MCND+04,key-2,key-11,MYDL04,YUNG2,KARBACH,key-1,TJO04,key-3,key-17,key-12,key-15,key-16,key-18,BBG,MAURO,key-6,PLENIO04,key-30}.
The main spirit of these articles is that permanently coupled systems
can be used for the transfer of quantum information with \emph{minimal
control,} that is, only the sending and the receiving parties can
apply gates to the system, but the part of the chain interconnecting
them cannot be controlled during the communication process. A scheme
with less control is obviously impossible. The first proposals \cite{SB03,key-1}
considered a regular spin chain with Heisenberg interactions. A physical
implementation of this scheme was discussed in \cite{JOSEPH}, and
its channel capacity was derived in \cite{key-7}. Already in \cite{SB03,key-1}
it was realized that such a transfer will, in general, not be perfect.
The reason for the imperfect transfer is the \emph{dispersion} of
the information along the chain. This becomes worse as the chains
get longer.

Since then, many very interesting methods have been developed to improve
the fidelity of the transfer. One method is to use Hamiltonians with
engineered couplings \cite{MCND+04,key-2,key-11,MYDL04,YUNG2,KARBACH} such that
the dispersed information will {}``refocus'' at the receiving end
of the chain. Another approach is to encode and decode the information
using multiple spins \cite{TJO04,key-3} to form Gaussian wave packets
(which have a lower dispersion). This has been generalized \cite{key-3}
in an elegant way by using {}``phantom'' spins such that a multiple-spin
encoding can be achieved by only controlling two sending and receiving
qubits. By using gapped systems \cite{key-17,key-12,key-15,key-16},
the intermediate spins are only virtually excited, and the transfer
has a very high fidelity. Finally, in \cite{key-18} we have suggested
a {}``dual-rail'' encoding using \emph{two parallel} quantum channels,
achieving and arbitrarily perfect transfer. It was shown in \cite{BBG}
that such a protocol achieves arbitrarily perfect transfer for nearly
any type of quantum chain, transforming a heavily dispersive dynamic
into one that can be used for state transfer. In this scheme, not
only the control needed during the transfer is minimized (no local
access along the chains is needed), but also the control needed to
\emph{design} the system in the first place. 

The main requirement for perfect transfer with dual-rail encoding
as presented in the literature till date \cite{key-18,BBG} is that
two \emph{identical} quantum chains have to be designed. While this
is not so much a theoretical problem, for possible experimental realizations
of the scheme \cite{JOSEPH} the question naturally arises how to
cope with slight asymmetries of the channels. The purpose of this
paper is to demonstrate that in many cases, perfect state transfer
with dual-rail encoding is possible for \emph{}quantum chains with
differing Hamiltonians. 

By doing so, we also offer a solution to another and perhaps more
\emph{general} problem: if one implements \emph{any} of the above
schemes, the Hamiltonians will always be different from the theoretical
ones by some random perturbation. This will lead to a decrease of
fidelity in particular where specific energy levels were assumed (see
\cite{key-19} for an analysis of fluctuations effecting the scheme
\cite{MCND+04}). Also in general, random systems can lead to a Anderson
localization \cite{key-29} of the eigenstates (and therefore to low
fidelity transport of quantum information). This problem can be avoided
using the scheme described below. We will show numerically that the
dual-rail scheme can still achieve arbitrarily perfect transfer for
a uniformly coupled Heisenberg with random noise on the coupling strengths
(both for the case of spatially correlated and uncorrelated fluctuations).
Moreover, for any two quantum chains, we show that Bob and Alice can
check whether their system is capable of dual-rail transfer without
directly measuring their Hamiltonians or local properties of the system
along the chains but by only measuring \emph{their} part of the system.

\section{Conclusive transfer}

\begin{figure}[htbp]
\begin{center}\includegraphics[%
  width=1.0\columnwidth]{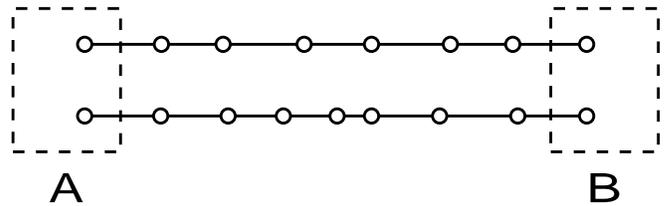}\end{center}

\caption{\label{fig:channel}Two quantum chains interconnecting $A$ and $B$.
Control of the systems is only possible at the two qubits of either
end.}
\end{figure}

By {}``conclusive transfer'', we mean that the receiver has a certain
probability for obtaining the {\it perfectly} transferred state, and some
way of checking whether this has happened. Conclusive transfer is more valuable than simple state transfer with the same fidelity, because errors are detected and memory effects \cite{WERNER} are unimportant if the transfer was successfull. A single spin-$1/2$ quantum chain could however
not be used for conclusive transfer, because any measurement can easily
destroy the unknown quantum state that is being transferred. The simplest
quantum chain for conclusive transfer is a system consisting of two
uncoupled quantum chains $(1)$ and $(2)$, as shown in Fig. \ref{fig:channel}.
The chains are described by the two Hamiltonians $H^{(1)}$ and $H^{(2)}$
acting on the corresponding Hilbert spaces $\mathcal{H}_{1}$ and
$\mathcal{H}_{2}.$ The total Hamiltonian is thus\begin{equation}
H=H^{(1)}\otimes I^{(2)}+I^{(1)}\otimes H^{(2)},\end{equation}
and the time evolution operator factorizes as\begin{eqnarray}
U(t) & = & \exp\left(-iHt\right)\\
 & = & \exp\left(-iH^{(1)}t\right)\otimes\exp\left(-iH^{(2)}t\right).\end{eqnarray}
For the moment, we assume that both chains have equal length $N$,
but it will become clear in Section \ref{sec:Tomography} that this
is not a requirement of our scheme. Like in \cite{key-18,BBG}, we
assume that the quantum chains consist of qubits and that both Hamiltonians
commute with the z-component of the total spin of chains,\begin{equation}
\left[H^{(1)},\sum_{i=1}^{N}\sigma_{z}^{(1)}\right]=\left[H^{(2)},\sum_{i=1}^{N}\sigma_{z}^{(2)}\right]=0.\end{equation}
 It then follows that the state\begin{equation}
\left|\mathbf{0}\right\rangle ^{(i)}\equiv\left|00\ldots0\right\rangle ^{(i)}\end{equation}
is an eigenstate of $H^{(i)}$ and that the dynamics of an initial
state of the form\begin{equation}
\left|\mathbf{1}\right\rangle ^{(i)}\equiv\left|10\ldots0\right\rangle ^{(i)}\end{equation}
is restricted to the subspace of single excitations, \begin{equation}
\left|\mathbf{m}\right\rangle ^{(i)}\equiv\left|0\ldots0\begin{array}[t]{c}
1\\
m\textrm{th}\end{array}0\ldots0\right\rangle ^{(i)}\quad(1\leq m\leq N).\end{equation}
In the following, we will omit all indexes and write the states of
the full Hilbert space $\mathcal{H}_{1}\otimes\mathcal{H}_{2}$ as\begin{equation}
\left|\mathbf{m,n}\right\rangle \equiv\left|\mathbf{m}\right\rangle ^{(1)}\otimes\left|\mathbf{n}\right\rangle ^{(2)}.\end{equation}
We assume that the sender, Alice, has full access to the first qubit
of each chain, and that the receiver, Bob, has full access to the
last qubit of each chain. With {}``full access'' we mean that they
can perform a two-qubit gate (say, a CNOT), and arbitrary single-qubit
operations. Bob also needs the ability to perform single-qubit measurements.%

\begin{figure}[htbp]
\begin{center}\includegraphics[%
  width=1.0\columnwidth]{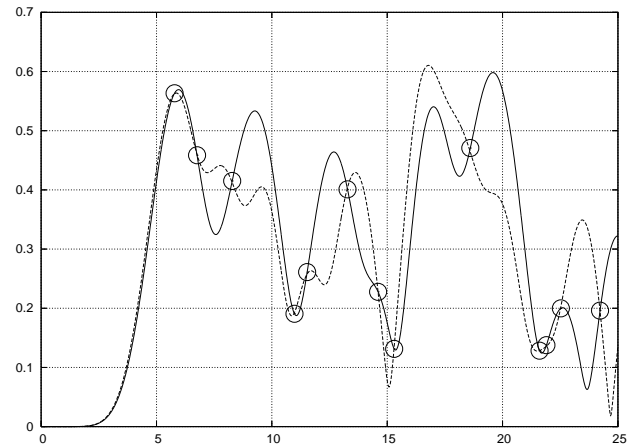}\end{center}

\caption{\label{fig:coincidence}The absolute values of the transition amplitudes
$f_{N,1}(t)$ and $g_{N,1}(t)$ for two Heisenberg chains of length
$N=10$. The couplings strengths of both chains were chosen randomly
from the interval $\left[0.8J,1.2J\right].$ The circles show times
where Bob can perform measurements without gaining information on
$\alpha$ and $\beta.$}
\end{figure}

Initially, Alice encodes the state as\begin{equation}
\alpha\left|\mathbf{0,1}\right\rangle +\beta\left|\mathbf{1,0}\right\rangle .\end{equation}
This is a superposition state of an excitation in the first qubit
of the first chain with an excitation in the first qubit of the second
chain. The state will evolve in\begin{equation}
\sum_{n=1}^{N}\left\{ \alpha g_{n,1}(t)\left|\mathbf{0,n}\right\rangle +\beta f_{n,1}(t)\left|\mathbf{n,0}\right\rangle \right\} .\label{eq:evolved}\end{equation}
Note that there is only one excitation in the system. The probability
amplitudes are given by \begin{eqnarray}
f_{n,1}(t) & \equiv & \left\langle \mathbf{n,0}\left|U(t)\right|\mathbf{1,0}\right\rangle \\
g_{n,1}(t) & \equiv & \left\langle \mathbf{0,n}\left|U(t)\right|\mathbf{0,1}\right\rangle .\end{eqnarray}
In \cite{key-18}, these functions were identical. For differing chains
this is no longer the case. We may, however, find a time $t_{1}$
such that the modulus of their amplitudes at the last spins are the
same (see Fig. \ref{fig:coincidence}),\begin{equation}
g_{N,1}(t_{1})=e^{i\phi_{1}}f_{N,1}(t_{1}).\label{eq:req1}\end{equation}
 At this time, the state (\ref{eq:evolved}) can be written as\begin{eqnarray}
\sum_{n=1}^{N-1}\left\{ \alpha g_{n,1}(t_{1})\left|\mathbf{0,n}\right\rangle +\beta f_{n,1}(t_{1})\left|\mathbf{n,0}\right\rangle \right\} +\nonumber \\
f_{N,1}(t_{1})\left\{ e^{i\phi_{1}}\alpha\left|\mathbf{0,N}\right\rangle +\beta\left|\mathbf{N,0}\right\rangle \right\} .\end{eqnarray}
Bob decodes the state by applying a CNOT gate on his two qubits, with
the first qubit as the control bit. The state thereafter is\begin{eqnarray}
\sum_{n=1}^{N-1}\left\{ \alpha g_{n,1}(t_{1})\left|\mathbf{0,n}\right\rangle +\beta f_{n,1}(t_{1})\left|\mathbf{n,0}\right\rangle \right\} +\nonumber \\
f_{N,1}(t_{1})\left\{ e^{i\phi_{1}}\alpha\left|\mathbf{0}\right\rangle ^{(1)}+\beta\left|\mathbf{N}\right\rangle ^{(1)}\right\} \otimes\left|\mathbf{N}\right\rangle ^{(2)}.\end{eqnarray}
Bob then measures his second qubit. Depending on the outcome of this
measurement, the systems will either be in the state\begin{equation}
\frac{1}{\sqrt{p_{1}}}\sum_{n=1}^{N-1}\left\{ \alpha g_{n,1}(t_{1})\left|\mathbf{0,n}\right\rangle +\beta f_{n,1}(t_{1})\left|\mathbf{n,0}\right\rangle \right\} \label{eq:failure}\end{equation}
or in\begin{equation}
\left\{ e^{i\phi_{1}}\alpha\left|\mathbf{0}\right\rangle ^{(1)}+\beta\left|\mathbf{N}\right\rangle ^{(1)}\right\} \otimes\left|\mathbf{N}\right\rangle ^{(2)},\label{eq:success}\end{equation}
where $p_{1}=1-\left|f_{N,1}(t_{1})\right|^{2}=1-\left|g_{N,1}(t_{1})\right|^{2}$
is the probability that Bob has \emph{not} received the state. The
state (\ref{eq:success}) corresponds to the correctly transferred
state with a known phase error (which can be corrected by Bob using
a simple phase gate). If Bob finds the system in the state (\ref{eq:failure}),
the transfer has been unsuccessful, but the information is still in
the chain. We thus see that conclusive transfer is still possible
with randomly coupled chains as long as the requirement (\ref{eq:req1})
is met. This requirement will be further discussed and generalized
in the next section.

\section{Arbitrarily perfect transfer}

If the transfer was unsuccessful, the state (\ref{eq:failure}) will
evolve further, offering Bob further opportunities to receive Alice's
message. For identical quantum chains, this must ultimately lead to
a success for any reasonable Hamiltonian \cite{BBG}. For differing
chains, this is not necessarily the case, because measurements are
only allowed at times where the probability amplitude at the end of
the chains are equal, and there may be systems where this is never
the case. In this section, we will develop a criterion that generalizes
Eq. (\ref{eq:req1}) and allows to check numerically whether a given
system is capable of arbitrarily perfect state transfer.

The quantity of interest for conclusive state transfer is the joint
probability $\mathcal{P}(l)$ that after having checked $l$ times,
Bob still has not received the proper state at his ends of the chain.
Optimally, this should approach zero if $l$ tends to infinity. In
order to derive an expression for $\mathcal{P}(l),$ let us assume
that the transfer has been unsuccessful for $l-1$ times with time
intervals $t_{i}$ between the the $i$th and the $(i-1)$th measurement,
and calculate the probability of failure at the $l$th measurement.
In a similar manner, we assume that all the $l-1$ measurements have
met the requirement of conclusive transfer (that is, Bob's measurements
are unbiased with respect to $\alpha$ and $\beta$) and derive the
requirements for the $l$th measurement. 

To calculate the probability of failure for the $l$th measurement,
we need to take into account that Bob's measurements disturb the unitary
dynamics of the chain. If the state before a measurement with the
outcome {}``failure'' is $\left|\psi\right\rangle ,$ the state
after the measurement will be\begin{equation}
\frac{1}{\sqrt{p_{l}}}Q\left|\psi\right\rangle ,\end{equation}
where $Q$ is the projector\begin{equation}
Q=I-\left|\mathbf{0,N}\right\rangle \left\langle \mathbf{0,N}\right|-\left|\mathbf{N,0}\right\rangle \left\langle \mathbf{N,0}\right|,\end{equation}
and $p_{l}$ is the probability of failure at the $l$th measurement.
The dynamics of the chain is alternating between unitary and projective,
such that the state before the $l$th measurement is given by\begin{equation}
\frac{1}{\sqrt{\mathcal{P}(l-1)}}\prod_{i=1}^{l}\left\{ U(t_{i})Q\right\} \left\{ \alpha\left|\mathbf{1,0}\right\rangle +\beta\left|\mathbf{0,1}\right\rangle \right\} ,\label{eq:product}\end{equation}
where we have used that \begin{equation}
Q\left\{ \alpha\left|\mathbf{1,0}\right\rangle +\beta\left|\mathbf{0,1}\right\rangle \right\} =\alpha\left|\mathbf{1,0}\right\rangle +\beta\left|\mathbf{0,1}\right\rangle \end{equation}
 for the first factor $i=1$, and \begin{equation}
\mathcal{P}(l-1)=\prod_{i=1}^{l-1}p_{l}.\label{eq:joint_equal_prod}\end{equation}
Note that the operators in (\ref{eq:product}) do not commute and
that the time ordering of the product (the index $i$ increases from
right to left) is important. The probability that there is an excitation
at the $N$th site of either chain is given by\begin{equation}
\frac{1}{\mathcal{P}(l-1)}\left\{ \left|\alpha\right|^{2}\left|F(l)\right|^{2}+\left|\beta\right|^{2}\left|G(l)\right|^{2}\right\} ,\end{equation}
with \begin{equation}
F(l)\equiv\left\langle \mathbf{N,0}\right|\prod_{i=1}^{l}\left\{ U(t_{i})Q\right\} \left|\mathbf{1,0}\right\rangle ,\label{eq:onlything}\end{equation}
and \begin{equation}
G(l)\equiv\left\langle \mathbf{0,N}\right|\prod_{i=1}^{l}\left\{ U(t_{i})Q\right\} \left|\mathbf{0,1}\right\rangle .\label{eq:onlything2}\end{equation}
Bob's measurements are therefore unbiased with respect to $\alpha$
and $\beta$ if and only if\begin{equation}
\left|F(l)\right|=\left|G(l)\right|\quad\forall l.\label{eq:condition}\end{equation}
In this case, the state can still be transferred conclusively (up
to a known phase). The probability of failure at the $l$th measurement
is given by\begin{equation}
p_{l}=1-\frac{\left|F(l)\right|^{2}}{\mathcal{P}(l-1)}.\label{eq:prob_fail_single}\end{equation}
We will show in the Appendix that the condition (\ref{eq:condition})
is equivalent to\begin{equation}
\left\Vert \prod_{i=1}^{l}\left\{ U(t_{i})Q\right\} \left|\mathbf{1,0}\right\rangle \right\Vert =\left\Vert \prod_{i=1}^{l}\left\{ U(t_{i})Q\right\} \left|\mathbf{0,1}\right\rangle \right\Vert \quad\forall l,\label{eq:finalcondition}\end{equation}
and that the joint probability of failure is simply\begin{equation}
\mathcal{P}(l)=\left\Vert \prod_{i=1}^{l+1}\left\{ U(t_{i})Q\right\} \left|\mathbf{1,0}\right\rangle \right\Vert ^{2}.\label{eq:finalprob}\end{equation}
 It may look as if Eq. (\ref{eq:finalcondition}) was a complicated
multi-time condition for the measuring times $t_{i}$, that becomes
increasingly difficult to fulfill with a growing number of measurements.
This is not the case. If proper measuring times have been found for
the first $l-1$ measurements, a trivial time $t_{l}$ that fulfills
Eq. (\ref{eq:finalcondition}) is $t_{l}=0.$ In this case, Bob measures
immediately after the $(l-1)$th measurement and the probability amplitudes
on his ends of the chains will be equal - but zero (a useless measurement).
But since the left and right hand side of Eq. (\ref{eq:finalcondition})
when seen as functions of $t_{l}$ are both quasi-periodic functions
with initial value zero, it is likely that they intersect many times,
unless the system has some specific symmetry or the systems are completely
different. Furthermore, for the $l$th measurement, Eq. (\ref{eq:finalcondition})
is equivalent to \begin{equation}
\left\Vert QU(t_{l})\left|\xi\right\rangle \right\Vert \equiv\left\Vert QU(t_{l})\left|\zeta\right\rangle \right\Vert \end{equation}
with \begin{equation}\left|\xi\right\rangle =Q\prod_{i=1}^{l-1}U(t_{i})Q\left|\mathbf{1,0}\right\rangle \end{equation}
and \begin{equation}\left|\zeta\right\rangle =Q\prod_{i=1}^{l-1}U(t_{i})Q\left|\mathbf{0,1}\right\rangle .\end{equation}
From this we can see that if the system is ergodic, the condition
for conclusive transfer is fulfilled at many different times.

Note that we do not claim at this point that any pair of chains will
be capable of arbitrary perfect transfer. We will discuss in the next
system how one can check this for a given system by performing some
simple experimental tests.

\section{Tomography\label{sec:Tomography}}

\begin{figure}
\includegraphics[%
  width=1.0\columnwidth]{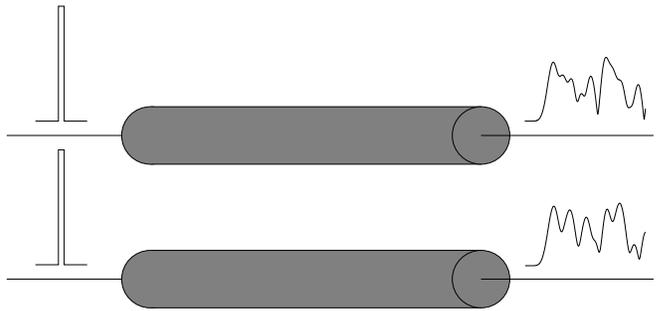}

\caption{\label{cap:blackbox}The relevant properties for conclusive transfer
can be determined by measuring the response of the two systems at
their ends only.}
\end{figure}

Suppose someone gives you two different experimentally designed spin
chains. It may seem from the above that knowledge of the full Hamiltonian
of both chains is necessary to check how well the system can be used
for state transfer. This would be a very difficult task, because we
would need access to all the spins along the channel to measure all
the parameters of the Hamiltonian. In fact by expanding the projectors
in Eq. (\ref{eq:finalcondition}) one can easily see that the only
matrix elements of the evolution operator which are relevant for conclusive
transfer are\begin{eqnarray}
f_{N,1}(t) & = & \left\langle \mathbf{N,0}\right|U(t)\left|\mathbf{1,0}\right\rangle \label{eq:t1}\\
f_{N,N}(t) & = & \left\langle \mathbf{N,0}\right|U(t)\left|\mathbf{N,0}\right\rangle \label{eq:t2}\\
g_{N,1}(t) & = & \left\langle \mathbf{0,N}\right|U(t)\left|\mathbf{0,1}\right\rangle \label{eq:t3}\\
g_{N,N}(t) & = & \left\langle \mathbf{0,N}\right|U(t)\left|\mathbf{0,N}\right\rangle .\label{eq:t4}\end{eqnarray}
Physically, this means that the only relevant properties of the system
are the transition amplitudes to \emph{arrive} at Bob's ends and to
\emph{stay} there. The modulus of $f_{N,1}(t)$ and $f_{N,N}(t)$
can be measured by initialising the system in the states $\left|\mathbf{1,0}\right\rangle $
and $\left|\mathbf{N,0}\right\rangle $ and then performing a reduced
density matrix tomography at Bob's site at different times $t$, and
the complex phase of these functions is obtained by initialising the
system in $\left(\left|\mathbf{0,0}\right\rangle +\left|\mathbf{1,0}\right\rangle \right)/\sqrt{2}$
and $\left(\left|\mathbf{0,0}\right\rangle +\left|\mathbf{N,0}\right\rangle \right)/\sqrt{2}$
instead. In the same way, $g_{N,1}(t)$ and $g_{N,N}(t)$ are obtained.
All this can be done in the spirit of \emph{minimal control} at the
sending and receiving ends of the chain only, and needs to be done
only once. It is interesting to note that the dynamics in the middle
part of the chain is not relevant at all. It is a {}``black box''
that may involve even completely different interactions, number of
spins, etc. (see Fig. \ref{cap:blackbox}). %
Once the transition amplitudes (Equations (\ref{eq:t1})-(\ref{eq:t4}))
are known, one can search numerically for optimized measurement times
$t_{i}$ using Eq. (\ref{eq:finalprob}) and the condition from Eq.
(\ref{eq:finalcondition}).

One weakness of the scheme described here is that the times at which
Bob measures have to be very precise, because otherwise the measurements
will not be unbiased with respect to $\alpha$ and $\beta.$ This
demand can be relaxed by measuring at times where not only the probability
amplitudes are similar, but also their \emph{slope} (see Fig. \ref{fig:coincidence}).
The computation of these optimal timings for a given system may be
complicated, but they only need to be done once.

\section{Numerical Examples}

In this section, we show some numerical examples for two chains with
Heisenberg couplings $J$ which are fluctuating. The Hamiltonians
of the chains are given by\begin{eqnarray}
H^{(1)} & = & \sum_{n=1}^{N-1}J(1+\delta_{n}^{(1)})\vec{\sigma}_{n}^{(1)}\cdot\vec{\sigma}_{n+1}^{(1)},\\
H^{(2)} & = & \sum_{n=1}^{N-1}J(1+\delta_{n}^{(2)})\vec{\sigma}_{n}^{(2)}\cdot\vec{\sigma}_{n+1}^{(2)},\end{eqnarray}
where $\delta_{n}^{(i)}$ uniformly distributed random numbers from
the interval $\left[-\Delta,\Delta\right].$ We have considered two
different cases: in the first case, the $\delta_{n}^{(i)}$ are completey
uncorrelated (i.e. independent for both chains and all sites along
the chain). In the second case, we have taken into account a spacial
correlation of the signs of the $\delta_{n}^{(i)}$ along each of
the chains, while still keeping the two chains uncorrelated. For both
cases, we find that arbitrarily perfect transfer remains possible
except for some very rare realisations of the $\delta_{n}^{(i)}.$ 

Because measurements must only be taken at times which fulfill the
condition (\ref{eq:finalcondition}) and these times usually do not
coincidence with the optimal probability of finding an excitation
at the ends of the chains, it is clear that the probability of failure
at each measurement will in average be higher than for chains without
fluctuations. Therefore, a bigger number of measurements have to be
performed in order to achieve the same probability of success. The
price for noisy couplings is thus a longer transmission time and a
higher number of gating operations at the receiving end of the chains.
Some averaged values are given in Table \ref{cap:The-total-time}.%
\begin{table}
\begin{center}\begin{tabular}{|c|c|c|c|c|c|}
\hline 
&
$\Delta=0$&
$\Delta=0.01$&
$\Delta=0.03$&
$\Delta=0.05$&
$\Delta=0.1$\tabularnewline
\hline
\hline 
$t$$\left[\frac{\hbar}{J}\right]$&
$377$&
$524\pm27$&
$694\pm32$&
$775\pm40$&
$1106\pm248$\tabularnewline
\hline 
$M$&
$28$&
$43\pm3$&
$58\pm3$&
$65\pm4$&
$110\pm25$\tabularnewline
\hline
\end{tabular}\end{center}

\caption{\label{cap:The-total-time}The total time $t$ and the number of
measurements $M$ needed to achieve a probability of success of $99$\%
for different fluctuation strengths $\Delta$ (uncorrelated case).
Given is the statistical mean and the standard deviation. The length
of the chain is $N=20$ and the number of random samples is $10.$
For strong fluctuations $\Delta=0.1$, we also found particular samples
where the success probability could not be achieved within the time
range searched by the algorithm.}
\end{table}
 for the Heisenberg chain with uncorrelated coupling fluctuations.

For the case were the signs of the $\delta_{n}^{(i)}$ are correlated,
we have used the same model as in \cite{key-19}, introducing the
parameter $c$ such that\begin{equation}
\delta_{n}^{(i)}\delta_{n-1}^{(i)}>0\qquad\textrm{with propability }c,\label{eq:corr1}\end{equation}
and \begin{equation}
\delta_{n}^{(i)}\delta_{n-1}^{(i)}<0\qquad\textrm{with propability }1-c.\label{eq:corr2}\end{equation}
For $c=1$ ($c=0)$ this corresponds to the case where the signs of
the couplings are completely correlated (anticorrelated). For $c=0.5$
one recovers the case of uncorrelated couplings. We can see from the
numerical results in Table \ref{cap:correlated} that arbitrarily
perfect transfer is possible for the whole range of $c.$ %
\begin{table}
\begin{center}\begin{tabular}{|c|c|c|c|c|c|c|}
\hline 
&
$c=0$&
$c=0.1$&
$c=0.3$&
$c=0.7$&
$c=0.9$&
$c=1$\tabularnewline
\hline
\hline 
$t$$\left[\frac{\hbar}{J}\right]$&
$666\pm20$&
$725\pm32$&
$755\pm41$&
$797\pm35$&
$882\pm83$&
$714\pm41$\tabularnewline
\hline 
$M$&
$256\pm2$&
$62\pm3$&
$65\pm4$&
$67\pm4$&
$77\pm7$&
$60\pm4$\tabularnewline
\hline
\end{tabular}\end{center}

\caption{\label{cap:correlated}The total time $t$ and the number of measurements
$M$ needed to achieve a probability of success of $99$\% for different
correlations $c$ between the couplings (see Eq. (\ref{eq:corr1})
and Eq. (\ref{eq:corr2})). Given is the statistical mean and the
standard deviation for a fluctuation strength of $\Delta=0.05$. The
length of the chain is $N=20$ and the number of random samples is
$20.$ }
\end{table}

\begin{figure}
\begin{center}\includegraphics[%
  width=1.0\columnwidth]{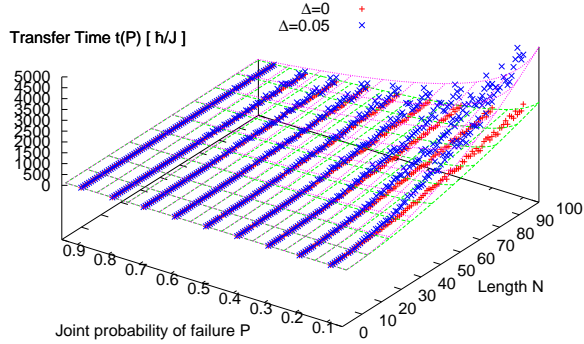}\end{center}

\caption{\label{cap:fitfig}Time $t$ needed to transfer a state with a given
joint probability of failure $P$ across a chain of length $N$ with
uncorrelated fluctuations of $\Delta=0.05$ and without fluctuations
($\Delta=0$ ). The points denote exact numerical data, and the fits
are given by Eq. (\ref{eq:fit}) and Eq. (\ref{eq:fit2}).}
\end{figure}

For $\Delta=0$, we know from \cite{key-18} that the time to transfer
a state with probability of failure $P$ scales as\begin{equation}
t(P)=\frac{0.33\hbar N^{1.6}}{J}\left|\ln P\right|.\label{eq:fit}\end{equation}
If we want obtain a similar formula in the presence of noise, we have
can perform a fit to the exact numerical data. For uncorrelated fluctuations
of $\Delta=0.05,$ this is shown in Fig. \ref{cap:fitfig}. The best
fit is given by\begin{equation}
t(P)=\frac{0.2\hbar N^{1.9}}{J}\left|\ln P\right|.\label{eq:fit2}\end{equation}
We conclude that weak fluctuations (say up to $5$\%) in the coupling
strengths do not deteriorate the performance of our scheme much for
the chain lengths considered. Both the transmission time and the number
of measurements raise, but still in a reasonable way (cf. Table \ref{cap:The-total-time}
and Eq. (\ref{cap:fitfig})). For larger fluctuations, the scheme
is still applicable in principle, but the amount of junk (i.e. chains
not capable of arbitrary perfect transfer) may get too large.%

\section{Conclusions}

We have shown that in many cases, it is possible to perfectly transfer
an unknown quantum state along a pair of quantum chains even if their
coupling is to some amount random. This is achieved using a dual-rail
encoding combined with measurements at the receiving end of the chains.
Since any scheme for quantum communication will suffer from some imperfections
when implemented, the dual-rail is a powerful tool to overcome the
decrease of fidelity.

This work was supported by the UK Engineering and Physical Sciences
Research Council through the grant GR/S62796/01 and the QIPIRC.

\appendix

\section*{Appendix}

We first use Eq. (\ref{eq:joint_equal_prod}) and (\ref{eq:prob_fail_single})
to obtain\begin{equation}
\mathcal{P}(l)=\mathcal{P}(l-1)-\left|F(l)\right|^{2},\end{equation}
or, using $\mathcal{P}(0)=1$,\begin{eqnarray}
\mathcal{P}(l) & = & 1-\sum_{k=1}^{l}\left|F(k)\right|^{2}.\end{eqnarray}
If we introduce\begin{eqnarray}
v(l) & \equiv & \left\Vert \prod_{i=1}^{l}\left\{ U(t_{i})Q\right\} \left|\mathbf{1,0}\right\rangle \right\Vert ^{2},\\
w(l) & \equiv & \left\Vert \prod_{i=1}^{l}\left\{ U(t_{i})Q\right\} \left|\mathbf{0,1}\right\rangle \right\Vert ^{2},\end{eqnarray}
then we can write \begin{eqnarray}
\lefteqn{\left|F(l)\right|^{2}=}\nonumber \\
 & = & \left\langle \mathbf{1,0}\right|\left(\prod_{i=1}^{l}\left\{ U(t_{i})Q\right\} \right)^{\dagger}\left|\mathbf{N,0}\right\rangle \left\langle \mathbf{N,0}\right|\prod_{i=1}^{l}\left\{ U(t_{i})Q\right\} \left|\mathbf{1,0}\right\rangle \nonumber \\
 & = & v(l)-\left\langle \mathbf{1,0}\right|\left(\prod_{i=1}^{l}\left\{ U(t_{i})Q\right\} \right)^{\dagger}Q\prod_{i=1}^{l}\left\{ U(t_{i})Q\right\} \left|\mathbf{1,0}\right\rangle \nonumber \\
 & = & v(l)-\left\Vert \prod_{i=1}^{l}Q\left\{ U(t_{i})Q\right\} \left|\mathbf{1,0}\right\rangle \right\Vert ^{2}\nonumber \\
 & = & v(l)-v(l+1),\end{eqnarray}
where we have used\begin{eqnarray}
\left|\mathbf{N,0}\right\rangle \left\langle \mathbf{N,0}\right| & = & 1-Q-\left|\mathbf{0,N}\right\rangle \left\langle \mathbf{0,N}\right|,\\
Q^{2} & = & Q,\end{eqnarray}
and the fact that the unitary matrix $U(t_{l+1})$ does not change
the norm of a vector. The same calculation can be done for $\left|G(l)\right|^{2}$
such that\begin{eqnarray}
\left|G(l)\right|^{2} & = & w(l)-w(l+1).\end{eqnarray}
If we want to have $\left|F\right|$ and $\left|G\right|$ matching
for all $l$, we need to have \begin{equation}
w(l)=v(l)\quad\forall l,\end{equation}
 or\begin{equation}
\left\Vert \prod_{i=1}^{l}\left\{ U(t_{i})Q\right\} \left|\mathbf{1,0}\right\rangle \right\Vert =\left\Vert \prod_{i=1}^{l}\left\{ U(t_{i})Q\right\} \left|\mathbf{0,1}\right\rangle \right\Vert \quad\forall l.\end{equation}
The joint probability of failure is given by\begin{eqnarray}
\mathcal{P}(l) & = & 1-\sum_{k=1}^{l}\left|F(k)\right|^{2}\\
 & = & 1-\sum_{k=1}^{l}v(k)-v(k+1)\\
 & = & v(l+1)\\
 & = & \left\Vert \prod_{i=1}^{l+1}Q\left\{ U(t_{i})Q\right\} \left|\mathbf{1,0}\right\rangle \right\Vert ^{2}.\end{eqnarray}

\end{document}